\documentclass[12pt]{article}

\usepackage{cite}
\usepackage{mathrsfs}

\usepackage{epsfig}

\usepackage{amsmath, amsthm, amssymb}

\usepackage{color}

\numberwithin{equation}{section}

\newcommand{\be}{\begin{equation}}
\newcommand{\ee}{\end{equation}}
\newcommand{\beq}{\begin{equation}}
\newcommand{\eeq}{\end{equation}}
\newcommand{\bea}{\begin{eqnarray}}
\newcommand{\eea}{\end{eqnarray}}
\newcommand{\gsim}{\lower.7ex\hbox{$\;\stackrel{\textstyle>}{\sim}\;$}}
\newcommand{\lsim}{\lower.7ex\hbox{$\;\stackrel{\textstyle<}{\sim}\;$}}

\catcode`@=11
\long\def\@caption#1[#2]#3{\par\addcontentsline{\csname
  ext@#1\endcsname}{#1}{\protect\numberline{\csname
  the#1\endcsname}{\ignorespaces #2}}\begingroup
    \small
    \@parboxrestore
    \@makecaption{\csname fnum@#1\endcsname}{\ignorespaces #3}\par
  \endgroup}
\catcode`@=12

\begin{document}

\vspace{-0.5cm}
\thispagestyle{empty}
\begin{flushright}
{\small IFT-UAM/CSIC-08-81}
 \end{flushright}

\vspace*{1.0cm}

\begin{center}
{\Large\textbf{Bayesian approach and Naturalness in MSSM analyses for the LHC}} 
\vspace*{0.8cm}

\textbf{M. E. Cabrera} $^a$\footnote{maria.cabrera@uam.es},
\textbf{J. A. Casas} $^a$\footnote{alberto.casas@uam.es} and
\textbf{R. Ruiz de Austri} $^b$\footnote{rruiz@ific.uv.es} 

\vspace{0.5cm}

\textit{$^a$ Instituto de F\'isica Te\'orica, IFT-UAM/CSIC, \\
U.A.M., Cantoblanco, 28049 Madrid, Spain}

\textit{$^b$ Instituto de F\'isica Corpuscular, IFIC-UV/CSIC, Valencia, Spain}

\end{center}

\vspace{0.5cm}
\begin{center}
{\bf Abstract}
\end{center}

The start of LHC has motivated an effort to determine the relative probability of the different regions of the MSSM parameter space, taking into account the present, theoretical and experimental, wisdom about the model. Since the present experimental data are not powerful enough to select a small region of the MSSM parameter space, the choice of a judicious prior probability for the parameters becomes most relevant. Previous studies have proposed theoretical priors that incorporate some (conventional) measure of the fine-tuning, to penalize unnatural possibilities. However, we show that such penalization arises from the Bayesian analysis itself (with no ad hoc assumptions), upon the marginalization of the $\mu-$parameter. Furthermore the resulting effective prior contains precisely the Barbieri-Giudice measure, which is very satisfactory. On the other hand we carry on a rigorous treatment of the Yukawa couplings, showing in particular that the usual practice of taking the Yukawas ``as required", approximately corresponds to taking logarithmically flat priors in the Yukawa couplings. Finally, we use an efficient set of variables to scan the MSSM parameter space, trading in particular $B$ by $\tan\beta$, giving the effective prior in the new parameters. Beside the numerical results, we give accurate analytic expressions for the effective priors in all cases. Whatever experimental information one may use in the future, it is to be weighted by the Bayesian factors worked out here.

\newpage
\setcounter{page}{1}

\section{Introduction}

The imminent start of LHC has motivated an interesting effort (see refs.~\cite{Allanach:2005kz, Allanach:2006jc, deAustri:2006pe, Allanach:2007qk, Roszkowski:2007fd, Buchmueller:2008qe, Trotta:2008bp, Ellis:2008di}) to anticipate which kind of supersymmetric model is more likely to be there, or, in more precise words, 
which region of the parameter space of the minimal supersymmetric standard model (MSSM) is more probable, taking into account the present (theoretical and experimental) wisdom about the model. This wisdom includes theoretical constraints (and perhaps prejudices) and experimental constraints, such as electroweak precision tests. The idea is to use this information to determine the relative probability of the different regions of the MSSM parameter space, thus the frequent expression ``LHC forecasts". The appropriate framework to evaluate this probability is the Bayesian approach, which allows to separate in a neat way the objective and subjective pieces of information. 

In the Bayesian analysis one tries to make inferences about the relative probability of different "states of nature" (corresponding to different values of the parameters defining the model, say $p_i$) upon the observation of different data which are determined\footnote{Normally this determination takes the form of a probability distribution since the theoretical computations and the experimental data are affected by different kinds of errors and uncertainties.} completely by $p_i$. 

The probability density of a particular point $\{p_i^0\}$ in the parameter space, given a certain set of {\em data}, is the so-called posterior probability density function (pdf), $p(p_i^0|{\rm data})$, which is given by the fundamental Bayesian relation (for a review see ref.~\cite{Trotta:2008qt})
\bea
\label{Bayes}
p(p_i^0|{\rm data})\ =\ p({\rm data}|p_i^0)\ p(p_i^0)\ \frac{1}{p({\rm data})}\ .
\eea
Here $p({\rm data}|p_i^0)$ is the likelihood (sometimes denoted by ${\cal L}$), i.e. the probability density of measuring the given data for the chosen point in the parameter space. E.g. for observables measured within a gaussian uncertainty, ${\cal L}$
is proportional to $e^{-\frac{1}{2}\chi^2}$, where $\chi^2$ is the conventional chi-squared. $p(p_i^0)$ is the prior, i.e. the ``theoretical" probability density that we assign a priory to the point in the parameter space. Finally, $p({\rm data})$ is a normalization factor which plays no role unless one wishes to compare different classes of models, so for the moment
it can be dropped from the previous formula. 

One can say that in eq.~(\ref{Bayes}) the first factor (the likelihood) is objective, while the second (the prior) is subjective, since it contains our prejudices about which regions of the parameter space are more ``natural" or ``expectable". It is desirable that the results of the analysis are as independent as possible of the chosen prior. This happens if the data are powerful enough to select a very small region of the parameter space, so that eq.~(\ref{Bayes}) is dominated by the likelihood, i.e. essentially the pdf is non-zero just in the narrow region of non-vanishing $p({\rm data}|p_i)$. However, in many instances this is not the case, as it happens for the MSSM. 

The somewhat subjective character of the prior, $p(p_i^0)$, has often motivated to ignore its presence, identifying in practice $p(p_i^0|{\rm data})$ with $p({\rm data}|p_i^0)$. However, it must be noticed that this procedure implicitly implies a choice for the prior, namely 
a completely flat prior in the parameters. This is not necessarily the most reasonable or "free of prejudices" attitude. Note for example that using $p_i^2$ as initial parameters 
instead of $p_i$ the previous flat prior becomes non-flat. So one needs some theoretical basis to establish, at least, the parameters whose prior can be reasonably taken as flat.

If we are interested in the most probable value of one (or several) of the initial parameters, say $p_i,\ i=1,...,N_1$, but not in the others, $p_i,\ i=N_1+1,...,N$, we have to {\em marginalize} the latter, i.e. integrate in the parameter space:
\bea
\label{marg}
p(p_i,\ i=1,...,N_1|{\rm data})\ = \int dp_{N_1+1},...,dp_N\ p(p_i,\ i=1,...,N|{\rm data})\ 
 .
\eea
This procedure is very useful and common to make predictions about the values of particularly interesting parameters. It must be noticed that, in order to perform the marginalisation, we need an input for the prior functions {\em and} for the range of allowed values of the parameters, which determines the range of the definite integration (\ref{marg}). A choice for these ingredients is therefore inescapable in trying to make LHC forecasts.

Let us now particularize these general statements to the MSSM (for a review see \cite{Martin:1997ns}). Beside the Standard Model (SM) --like parameters (to be discussed below), the MSSM contains a great number of parameters associated with the unknown process of supersymmetry (SUSY) breaking, the so-called soft SUSY-breaking terms. Assuming universality of these terms at a given high scale (namely the scale at which the SUSY breaking is transmitted to the observable sector), these parameters are reduced to four: the universal scalar mass, $m$, the universal gaugino mass, $M$, the universal trilinear scalar coupling, $A$, and the bilinear scalar coupling, $B$. The universality assumption is in part justified by the need of keeping the FCNC processes under control and it does come out naturally in several schemes of SUSY breaking mediation, e.g. minimal SUGRA or gauge-mediated models (for a review see \cite{Nilles:1983ge} and \cite{Kolda:1997wt} respectively). Beside these four parameters one has to include the $\mu$-parameter (i.e. the Higgs mass term in the superpotential) as an additional independent parameter, presumably with a magnitude similar to the soft breaking terms, as it is demanded by a successful electroweak breaking (see below). The notation used here is consistent with refs.~\cite{Martin:1997ns, softsusy}.

The SM-like parameters of the MSSM include the $SU(3)\times SU(2)\times U(1)_Y$ gauge couplings, $g_3,g,g'$, and the Yukawa couplings, which in turn determine the fermion masses and mixing angles. An important difference from the SM is that the MSSM contains two Higgs doublets, $H_1, H_2$, with expectation values
$v_i=\langle H_i^0\rangle$ determined by the parameters of the model upon minimization of the scalar potential, $V(H_1, H_2)$. They have to fulfill $2(v_1^2 + v_2^2) = v^2 = (246\ {\rm GeV})^2$.
The down-type-quark masses go like $m_d \sim y_d v_1= y_d v \cos \beta$, where $\tan \beta \equiv v_2/v_1$. Similarly for the up-type-quarks $m_u \sim y_u v_2= y_u v \sin \beta$, and for the charged leptons, $m_e \sim y_e v_1= y_e v \cos \beta$. Hence the values of the Yukawa couplings which give the observed fermion masses depend on the derived parameter $\tan\beta$, a fact that will be relevant later in our discussion.

In sect.~2 we address some basic aspects of the Bayesian approach for the MSSM, showing in particular that a penalization of the fine-tuning arises from the Bayesian analysis itself (with no ad hoc assumptions as in previous analyses), upon the marginalization of the $\mu-$parameter (subsect.~2.1). We also present a 
rigorous treatment of the Yukawa couplings, showing that the usual practice of taking the Yukawas ``as required", approximately corresponds to taking logarithmically flat priors in the Yukawa couplings (subsect.~2.2). In sect.~3
we use an efficient set of variables to scan the MSSM parameter space, trading in particular $B$ by $\tan\beta$, giving the effective prior in the new parameters. 
Finally, in sect.~4 we summarize our results and conclusions.

\section{Some basic aspects}

\subsection{Connection between the Bayesian approach and the fine-tuning measure}

It is common lore that the parameters of the MSSM, $\{m,M,A,B,\mu\}$, should not be far from the electroweak scale in order to avoid unnatural fine-tunings to obtain the correct scale of the electroweak breaking. This can be easily appreciated from the minimization of the tree-level form of the scalar potential, $V(H_1,H_2)$, which gives the expectation values of the Higgses, and thus the value of $M_Z^2= \frac{1}{2}(g^2+g'^2)(v_1^2+v_2^2)$; namely 
\bea
\label{MZ}
M_Z^2= 2\ \frac{m_{H_1}^2 - m_{H_2}^2\tan^2 \beta}{\tan^2\beta-1} -2\mu^2\ \ .
\eea
Unless the $\mu-$term and the soft masses $m_{H_i}$ (which upon the renormalization running depend also on the other soft terms) are close to the electroweak scale, a funny cancellation among the various terms in the right hand side of (\ref{MZ}) is necessary to get the experimental $M_Z$. 

A conventional measure of the degree of fine-tuning is given by the Barbieri-Giudice
fine-tuning parameters \cite{Barbieri:1987fn}:
\bea
\label{BG}
c_i = \left|\frac{\partial \ln M_Z^2}{\partial \ln p_i}\right|,\;\;\;\; 
\eea
which weigh up the sensitivity of $M_Z$ with respect to the parameters of the model, $p_i$. The global measure of the fine-tuning is taken as $c\equiv {\rm max}\{c_i\}$ or $c\equiv \sqrt{\sum c_i^2}$ \cite{Barbieri:1987fn, Ciafaloni:1996zh, Casas:2004gh, Casas:2005ev}.

Previous studies have attempted to incorporate this fine-tuning measure to the Bayesian approach through the prior $p(p_i)$. In particular, in refs.~\cite{Giusti:1998gz, Allanach:2006jc} a prior $p(p_i)\propto 1/c$ was proposed\footnote{Another prior designed to catch the naturalness criterion has been proposed in ref.~\cite{Allanach:2007qk}.}. In principle this is not unreasonable since $1/c$ approximately
indicates the probability of a cancellation among the various terms contributing to $M_Z^2$ to give a result $\lsim (M_Z^{\rm exp})^2$.
This can be intuitively seen as follows. Expanding $M_Z^2(p_i)$ around a point in parameter space that gives the desired cancellation, say ${\cal P}^0\equiv \{p_i^0\}$,
up to the linear term in the parameters, one finds that only a small neighborhood $\delta {\cal P}\sim {\cal P}^0/c$ around this point gives a value of $M_Z^2$ smaller or equal to the experimental value \cite{Ciafaloni:1996zh}. Hence, if one assumes that ${\cal P}$ could reasonably have taken any value of the order of magnitude of ${\cal P}^0$, then only for a small fraction $\sim 1/c$ of this region one gets $M_Z^2\lsim (M_Z^{\rm exp})^2$, thus the rough probabilistic meaning of $c$.
 
However, though reasonable, the above-mentioned proposals for priors are rather arbitrary, as the very measure of the fine-tuning is. On the other hand, since the naturalness arguments are deep down statistical arguments, one might expect that an effective penalization of fine-tunings
should arise from the Bayesian analysis itself, with no need of introducing 
"naturalness priors" {\em ad hoc}. This is in fact the case, as we are about to see.

Let us consider $M_Z$ as an experimental data, on a similar foot to the rest of physical observables. Then the total likelihood reads
\bea
\label{likelihood}
p({\rm data}|s, m, M, A, B, \mu)\ =\ N_Z\ e^{-\frac{1}{2}\chi_Z^2}\ {\cal L}_{\rm rest}\ ,
\eea
where $s$ represents the SM-like parameters,  ${\cal L}_{\rm rest}$ is the likelihood associated to all the physical observables, except $M_Z$,
and 
\bea
\label{likelihood_Z}
\chi_Z^2 = \left(\frac{M_Z-M_Z^{\rm exp}}{\sigma_Z}\right)^2\ ,
\eea
where $\sigma_Z\ll M_Z^{\rm exp}$ is the experimental uncertainty in the $Z$ mass;
finally $N_Z=1/\sqrt{2\pi\sigma_Z}$ is a normalization constant.
Let us now use this 
sharp dependence on $M_Z$ to marginalize the pdf in the $\mu-$parameter, performing a change of variable $\mu\rightarrow M_Z$:
\bea
\label{marg_mu}
p(s, m, M, A, B| \ {\rm data} )& = &\int d\mu\ p(s, m, M, A, B, \mu | 
{\rm data} )
\nonumber \\
&=& N_Z\ \int dM_Z \left[\frac{d\mu}{d M_Z}\right] e^{-\chi_Z^2}\ {\cal L}_{\rm rest}\ p(s, m, M, A, B, \mu)
\nonumber\\ 
&\simeq&\ {\cal L}_{\rm rest}  \left[\frac{d\mu}{d M_Z}\right]_{\mu_0}
p(s, m, M, A, B, \mu_0)\ .
\eea
where $\mu_0$ is the value of $\mu$ that reproduces the experimental value
of $M_Z$ for the given values of $\{s, m, M, A, B\}$. In the last line of (\ref{marg_mu}) we have approximated  $N_Z\ e^{-\frac{1}{2}\chi_Z^2}\simeq \delta(M_Z-M_Z^{\rm exp})$. Essentially the same result is obtained by performing the $\mu-$integration in the stationary point approximation. Now, comparing (\ref{marg_mu}) to the definition of fine-tuning parameters (\ref{BG}), we can write
\bea
\label{pcmu}
p(s, m, M, A, B| \ {\rm data} )\
=\ 2\ {\cal L}_{\rm rest}\  \frac{\mu_0}{M_Z}\ \frac{1}{c_\mu}\
p(s, m, M, A, B, \mu_0)\ .
\eea
Several comments are in order here. First, the presence of the fine-tuning parameter, $1/c_\mu$, penalizes the regions of the parameter space with large fine-tuning, as desired. Actually eq.~(\ref{pcmu}) is very similar to multiply by hand the initial prior in the parameters by a factor $1/c$, as in ref.~\cite{Allanach:2006jc}. The difference is that here the factor $1/c_\mu$ has not been put by hand: it comes out from the marginalization in $\mu$. Moreover the prior $p(s, m, M, A, B, \mu_0)$ is still undefined. If one takes it as flat, then one gets the same as in ref.~\cite{Allanach:2006jc}, but with one factor $\mu$ in the numerator (still the regions of large fine-tuning are penalized since $c_\mu$ goes parametrically as $\sim \mu^2$). If one takes logarithmically flat priors, i.e. $p(\mu)\propto 1/\mu$, then eq.~(\ref{pcmu}) would formally coincide with the procedure of multiplying the theoretical prior $p(s, m, M, A, B)$ by a factor $1/c$.
This is reasonable: the usual naturalness criteria implicitly assume that for a given value of one parameter, say $\mu=\mu_0$, the prior probability is distributed around $\mu_0$ \cite{Ciafaloni:1996zh, Casas:2005ev} with a width $\sim \mu_0$ [see the brief discussion in the paragraph after eq.~(\ref{BG})]. This is equivalent to assume that the value $\mu=\mu_0$ has a prior probability $\propto 1/\mu_0$.
Actually this is the reason why, according to usual fine-tuning arguments, large soft parameters are more unlikely than small ones: for the former the region of the parameter space that produces the observed electroweak scale is much narrower than for the latter,
{\em not} in absolute value, but {\em compared} to the size of the soft parameters in each case. Assuming flat priors there would be no reason to prefer soft parameters of the electroweak size instead of e.g. order $M_{\rm GUT}$. The fact that even for flat priors we still get a penalty factor $\mu/c_\mu$ comes from the assumption of a prior flat in $\mu$ instead of $\mu^2$, which is the quantity that appears in the cancellation [see e.g. eq.~(\ref{MZ})].

We find very satisfactory that the usual parameter to quantify the degree of fine-tuning emerges from the Bayesian approach ``spontaneously", not upon subjective assumptions, especially taking into account that there has been much discussion in the literature about its significance and suitability, see e.g. refs.~\cite{deCarlos:1993yy, Ciafaloni:1996zh, Casas:2004gh, Casas:2005ev, Giusti:1998gz}. Actually, one gets simply $c_\mu$ instead  $c$, as defined in eq.~(\ref{BG}). Of course there is nothing special with the $\mu-$parameter, except the fact that we have chosen to marginalize it using the experimental information about $M_Z$, which is the usual practice. Had we chosen to marginalize another parameter, say $M$, we would have got $c_M$, but of course at the end the results would be the same.

\begin{figure}
\centering
\includegraphics[angle=0,width=.46\linewidth]{ft-a0-b0-mu+-fp.eps}\;\;\;\;\;\;\;\;
\includegraphics[angle=0,width=.46\linewidth]{ft-a0-b0-mu+-logp.eps}\vspace{0.6cm}
\includegraphics[angle=0,width=.46\linewidth]{msugra-ft-a0-mu+-fp.eps}\;\;\;\;\;\;\;\;
\includegraphics[angle=0,width=.46\linewidth]{msugra-ft-a0-mu+-logp.eps}

\caption{Values of the factor $\mu\ p(\mu)/(M_Z c_\mu)$ (in logarithmic units and up to a convenient proportionality constant)
in the $\{m, M\}$ plane for $\mu>0, A=0,\ B=0$ (upper plots), and for $\mu<0,\ A=0$ and the minimal SUGRA relation $B=A-m$ (lower plots), using
the two basic initial priors, 
$p(\mu)\propto {\rm const.}$ (left plots), $\ p(\mu)\propto 1/\mu$ (right plots).
The plotted factor appears in the effective prior given in eq.~(\ref{c_factor}).
\label{fig:count}}
\end{figure}

A convenient way to view eq.~(\ref{pcmu}) is to imagine that we start with an 
MSSM parameter space $\{s, m, M, A, B\}$ where $\mu$ has been eliminated 
using the experimental value of $M_Z$. Then the pdf appears as the
likelihood associated to the experimental information (except $M_Z^{\rm exp}$)
times an effective prior
\bea
\label{c_factor}
p_{\rm eff}(s, m, M, A, B)\
=\ 2\ \frac{\mu_0}{M_Z}\ \frac{p(\mu_0)}{c_\mu}\
p(s, m, M, A, B)\ ,
\eea
where for simplicity we have assumed that the prior in $\mu$ factorizes from the rest.
This means that the initial prior gets multiplied by a factor 
$2\frac{\mu_0}{M_Z}\ \frac{p(\mu_0)}{c_\mu}$ that carries the fine-tuning penalty.
In Fig.~1 we have plotted this factor in representative slices of 
the $\{s, m, M, A, B\}$ 
parameter space (using the two basic choices $p(\mu)\propto {\rm const.},\ p(\mu)\propto 1/\mu$) for some illustrative and physically relevant cases. In all of them large soft parameters get penalized (except partially for focus-point regions \cite{Feng:1999mn, Feng:1999zg}). There are no ad hoc assumptions for this result, it just comes out from  
the value of $M_Z^{\rm exp}$ and the marginalization of $\mu$. 

For practical calculations it is useful to have an approximate expression for $c_\mu$. From the tree-level condition (\ref{MZ}) we see that $c_\mu\sim 2\mu^2/M_Z^2$. Nevertheless, using the approximate analytic formulas discussed in sect.~3, it is possible to write a much more refined expression for $c_\mu$, which we postpone to that section.

\subsection{Nuisance variables and the role of the Yukawa couplings}

It is common in statistical problems that not all the parameters that define the system are of interest. In the problem at hand we are interested in determining the probability regions 
for the MSSM parameters that describe the new physics, i.e. $\{m,M,A,B,\mu\}$, but not (or not at the same level) in the SM-like parameters, denoted by $\{s\}$. However, the {\em nuisance parameters} $\{s\}$ play an important role in extracting experimental consequences from the MSSM. The usual technique to eliminate nuisance parameters is simply marginalizing them, i.e. integrating the pdf (\ref{pcmu}) in the $\{s\}$ variables (for a review see ref.~\cite{Berger:1999}). When the value of a nuisance parameter is in one-to-one correspondence to a high-quality experimental piece of information (included in ${\cal L}_{\rm rest}$), this integration simply selects the ``experimental" value of the nuisance parameter, which thus becomes (basically) a constant with no further statistical significance in the analysis. In particular, the prior on such nuisance parameter becomes irrelevant. In the MSSM, nuisance parameters of this class are the gauge couplings, $\{g_3,g,g'\}$\footnote{Strictly speaking, the initial theoretical inputs are the gauge couplings at high energy, which are related to the experimental (low-energy) ones by the renormalization-group running. This running depends on the other MSSM parameters through the position of thresholds associated with different particles. Hence, two viable MSSM models have slightly different values of the gauge couplings at high energy, and thus the theoretical prior on the couplings would play an (almost insignificant) role in the statistical comparison of the two models.}, which thus can be extracted from the analysis. 


In the pure SM a similar argument can be used to eliminate the Yukawa couplings, since they  are in one-to-one correspondence to the quark and lepton masses. However, as discussed in sect.~1, in the MSSM these masses depend also on the value of $\tan\beta\equiv v_2/v_1$, which is a derived quantity that takes different values at different points of the MSSM parameter space. This means that two viable MSSM models (with the same fermion masses) will have in general very different values of the Yukawa couplings, and thus the theoretical prior, $p(y)$, will play a relevant and non-ignorable role in their relative probability. Any Bayesian analysis of the MSSM amounts to an explicit or implicit assumption about the prior in the Yukawa couplings.

In order to make these points more explicit, let us temporarily simplify the discussion approximating the experimental likelihood related to the fermion masses as
\bea
\label{likelihood_mt}
{\cal L}_{\rm fermion\ masses}\ =\ \delta(m_t-m_t^{\rm exp})\ \delta(m_b-m_b^{\rm exp})\ ....
\eea
(which is a fair approximation). This is a factor of the global likelihood, ${\cal L}_{\rm rest}$. Likewise, let us approximate the theoretical values of the fermion masses as
\bea
\label{mt}
m_t= \frac{1}{\sqrt{2}}y_t^{\rm low} v s_\beta, \;\;
m_b= \frac{1}{\sqrt{2}}y_b^{\rm low} v c_\beta,\;\;\;\;\;
\rm{etc.}
\eea
where $s_\beta\equiv \sin\beta$, $c_\beta\equiv \cos\beta$ and $y_i^{\rm low}$ are the low-energy Yukawa couplings. As it is well-known these expressions correspond to the running masses. The physical (pole) masses include a radiative correction that we have ignored here, but not in our full analysis. A further simplification is to assume $y_i^{\rm low} = R_i y_i$, where $y_i$ are the high-energy Yukawa couplings (and thus the input parameters) and the renormalization-group factor $R_i$ does not depend on $y_i$ itself (this is not a good approximation for the top Yukawa coupling, but we will assume it momentarily for the sake of clarity). Now, the marginalization in the Yukawa couplings can be readily done, integrating the pdf given by eq.~(\ref{pcmu}) in the $y_i$ variables.
Writing just the relevant terms we get
\bea
\label{marg_y}
&&\hspace{-1cm}\int [dy_t\ dy_b\cdots]\ p(y, m, M, A, B| \ {\rm data} )\
=\ \int [dy_t\ dy_b\cdots]\ p(y) 
\delta(m_t-m_t^{\rm exp})\ \delta(m_b-m_b^{\rm exp})\cdots
\nonumber\\
&&\hspace{3cm}\sim\  p(y)\ \left|\frac{d y_t}{d m_t}\right|\left|\frac{d y_b}{d m_b}\right|\cdots\ 
=\ p(y)\ s_\beta^{-1}\ c_\beta^{-1}\cdots
\eea
where $p(y)$ denotes the prior in the Yukawa couplings (which we assume that factorizes from the other priors). Eq.~(\ref{marg_y}) represents the footprint of the Yukawa couplings in the pdf.
Note that the factors $s_\beta^{-1}\ c_\beta^{-1} \cdots$ arise from the change of variables 
$y_i\rightarrow m_i$, even if the likelihood is not approximated by deltas. There are as many such factors as quarks and leptons. This amounts to a dramatic modulation of the relative probability of MSSM regions with different $\tan \beta$ if one chooses a flat prior, $p(y)=$ const. If, instead, one takes logarithmically flat priors, i.e. $p(y_i)\propto 1/y_i$, then the $s_\beta^{-1}\ c_\beta^{-1}\cdots$ factors get cancelled, so that the elimination of
the Yukawa couplings does not leave a footprint in the probability density of the (non-nuisance) MSSM parameter space, $\{m,M,A,B,\mu\}$. 

In previous Bayesian analyses of the MSSM the role of the Yukawa couplings was not considered to this extent. Essentially, their values were taken as needed to reproduce the experimental fermion masses, within uncertainties. As we have seen, this practice approximately corresponds to assuming logarithmically flat priors in the Yukawa couplings\footnote{Actually, for independent reasons, we find the logarithmically flat prior for Yukawa couplings a most sensible choice. Certainly there is no convincing origin for the experimental pattern of fermion masses, and thus of Yukawa couplings. However it is a fact that these come in very assorted orders of magnitude (from ${\cal O}(10 ^{-6})$ for the electron to ${\cal O}(1)$ for the top), suggesting that the underlying mechanism may
produce Yukawa couplings of different orders with similar efficiency.}. 

The above discussion is however oversimplified. As already mentioned, the marginalization in the top Yukawa coupling (and sometimes the bottom one) produces extra factors due to the dependence of $R_t$ on $y_t$. Actually, since one is marginalizing simultaneously in the Yukawa couplings and the $\mu-$parameter one has to evaluate the full Jacobian of the transformation $\{\mu, y_t\}\rightarrow \{M_Z, m_t\}$, which introduces additional contributions. Furthermore, the picture gets more complicated due to the fact that, for a given choice of $\{m,M,A,B\}$, there may be several values of
$\mu$ leading to the correct value of $M_Z$ with different values of $\tan\beta$ and thus of the Yukawa couplings. This means that in the marginalization one has to sum over all these possibilities. This is technically annoying and reduces the clarity of the approach. These drawbacks can be eliminated by trading in the statistical analysis the initial $B-$parameter by the derived $\tan\beta$ parameter, as we discuss in the next section.

Let us finally mention that in the analysis of ref.~\cite{Allanach:2006jc} the fermion masses themselves, rather than the Yukawa couplings, were taken as SM-like variables. The advantage of such procedure is that these nuisance variables are in obvious one-to-one correspondence to the experimental data. Then the priors on the masses become almost irrelevant, and they can be integrated out, almost without leaving any footprint. However, this has two problems. First, the fermion masses are obviously derived quantities and should not be taken as initial input variables, even if this makes life easier. Second, such procedure introduces completely artificial factors, as it will become clear at the end of the next section.

\section{Efficient variables to scan the MSSM parameter space}

In MSSM analyses it is normally very advantageous, both for theoretical and phenomenological reasons, to trade the initial $B-$parameter by the derived $\tan\beta$ parameter. On the phenomenological side, $\tan\beta$ is a parameter that appears explicitly in the predictions for many physical processes, such as cross sections, branching ratios, etc. (this is unlike  $B$, that enters only in a very indirect way). Thus it is convenient to get the probability density of the MSSM parameter space as a function of $\tan\beta$. On the theoretical side, for a given viable choice of $\{m,M,A,\tan\beta\}$, there are exactly two values of $\mu$ (with opposite sign and the same absolute value at low energy) leading to the correct value of $M_Z$. Thus working in one of the two (positive and negative) branches of $\mu$, each
point in the $\{m,M,A,\tan\beta\}$ space corresponds exactly to one model, whereas a point
in the $\{m,M,A,B\}$ space may correspond to several models, introducing a conceptual and technical complication in the analysis, as mentioned in the previous section.

Changing variables $B\rightarrow \tan\beta$ amounts to a factor $dB/d\tan\beta$ in
the pdf. On the other hand, we have seen in sect.~2  that it is convenient to trade $\mu$ and $y_t$ by $M_Z$ and $m_t$, as this makes the marginalization of these variables easier and more transparent. 
Thus we should compute the whole Jacobian, $J$, of the transformation
\bea
\label{change_3}
\{\mu,y_t,B\}\ \rightarrow\  \{M_Z,m_t,t\},\;\;\;\;\;\;t\equiv\tan\beta\ ,
\eea
so that, in the new variables, the pdf reads
\bea
\label{pdf_new}
p(g_i,m_t, m, M, A, \tan\beta| \ {\rm data} )\
= \ {\cal L}_{\rm rest}\ J|_{\mu=\mu_0}\ p(g_i, y_t, m, M, A, B, \mu=\mu_0)\ .
\eea
Here we have made explicit the dependence on the gauge couplings, and the top Yukawa coupling and mass, but not on the other fermions'. In this equation we have already marginalized $M_Z$ using the associated likelihood $\sim \delta(M_Z-M_Z^{\rm exp})$ (recall that $\mu_0$ is the value of $\mu$ that reproduces the experimental 
$M_Z$.)
The combination
\bea
\label{eff_prior}
p_{\rm eff}(g_i, m_t, m, M, A, \tan\beta)\  \equiv\ 
J|_{\mu=\mu_0}\  p(g_i, y_t, m, M, A, B, \mu=\mu_0)\
\eea
can be viewed as the effective prior in the new, more convenient, variables to scan the MSSM. Note that, as discussed in subsect.~2.2, the gauge couplings are fairly irrelevant for the statistical analysis, so we will drop them in what follows.
In order to work out $J$ we need the dependence of the old variables on the new ones, which can be derived from the minimization equations of the scalar potential, $V(H_1,H_2$), and from the expression of the top pole mass. For the numerical analysis we have used the SOFTSUSY code \cite{softsusy} which implements the full one-loop contributions and leading two-loop terms to the tadpoles for the electroweak symmetry breaking conditions with parameters running at two-loops. This essentially corresponds to the next-to-leading log approximation. 
However, in order to highlight
the most relevant facts it is useful to write down the expressions arising from the minimization of the tree-level potential with parameters running at one-loop (i.e. essentially the leading log approximation):
\bea
\label{mu}
\mu_{\rm low}^2= \frac{m_{H_1}^2 - m_{H_2}^2t^2}{t^2-1} - \frac{M_Z^2}{2}
\eea
\bea
\label{B}
B_{\rm low}= \frac{s_{2\beta}}{2\mu_{\rm low}}(m_{H_1}^2 + m_{H_2}^2+2\mu_{\rm low}^2)
\eea
\bea
\label{y}
y_{\rm low}=\frac{m_t}{v\ s_\beta}\ .
\eea
Here the ``low'' subscript indicates that the quantity is evaluated at low scale (more precisely, at a representative supersymmetric mass, such as the geometric average of the stop masses). The soft masses $m_{H_i}^2$ are also understood at low scale. For notational simplicity, we have dropped the subscript $t$ from the Yukawa coupling. 
We are not making explicit the role of the bottom Yukawa coupling, which is treated in a similar foot to the top one. Note that all these low-energy quantities contain an implicit dependence on the top Yukawa coupling through the corresponding renormalization-group equations (RGEs). 
The effect of the one-loop corrections on the effective potential to the previous expressions is incorporated by correcting the soft masses $m_{H_i}^2$ with one-loop tadpole effects along the lines of ref.~\cite{Pierce:1996zz}. Similarly the pole top mass is given by the running top mass, appearing in eq.~(\ref{y}), plus a radiative correction $\Delta_{rad} m_t$.
Eqs.~(\ref{mu}--\ref{y}), even when corrected with the mentioned radiative effects, have the structure
\bea
\label{muyB}
\mu=f(M_Z,y,t),\;\;\; y=g(M_Z,m_t,t),\;\;\; B=h(\mu,y,t)\ ,
\eea
where we only make explicit the dependence on the variables involved in the change of variables (\ref{change_3}). Note that $y$ depends on $M_Z$ since $v\propto M_Z$. Notice also that, unlike eqs.~(\ref{mu}--\ref{y}),  eqs.~(\ref{muyB}) are defined in terms of the the high-energy parameters. 

From eqs.~(\ref{muyB}) it is straightforward to evaluate the Jacobian $J$
of the transformation (\ref{change_3}), and thus the effective prior (\ref{eff_prior}).  $J$ gets simply
\bea
\label{J}
J = \left|\begin{array}{ccc} 
\frac{\partial\mu}{\partial M_Z}& \frac{\partial\mu}{\partial t} & \frac{\partial\mu}{\partial m_t}
\\
& & \\
\frac{\partial B}{\partial M_Z}& \frac{\partial B}{\partial t} &  \frac{\partial B}{\partial m_t} 
\\
& & \\
\frac{\partial y}{\partial M_Z}& \frac{\partial y}{\partial t} 
& \frac{\partial y}{\partial m_t} 
\end{array}\right|
\ =\ \frac{\partial f}{\partial M_Z}\ \frac{\partial g}{\partial m_t}
\ \frac{\partial h}{\partial t} \ \ ,
\eea
where the factor $\partial f/\partial M_Z$ carries essentially the fine-tuning penalization discussed in subsect.~2.1. 

We can give an analytical and quite accurate expression of $J$ by using the approximate equations (\ref{mu}--\ref{y}), and expressing the low-energy values of $\mu, B, y$ in terms of the 
high-energy ones through the integrated 1-loop RGEs. Schematically,
\bea
\label{muBLH}
\mu_{\rm low}= R_\mu(y)\mu,\;\;\;\; B_{\rm low}=B + \Delta_{RG}B(y)\ ,
\eea
where $R_\mu(y), \Delta_{RG}B(y)$ are definite functions of $y$ (and other parameters, but not $\mu$ and $B$) \cite{eg}. Similarly,
\bea
\label{yLH}
y_{\rm low}\simeq \frac{y E(Q_{\rm low})}{1+6yF(Q_{\rm low})}\ ,
\eea
where $Q$ is the renormalization scale, $F = \int_{Q_{\rm high}}^{Q_{\rm low}} E \ln Q$, and $E(Q)$ is a definite function that depends just on the gauge couplings \cite{Ibanez:1983di}. Plugging
(\ref{muBLH}) and (\ref{yLH}) into eqs.~(\ref{mu}--\ref{y}) we get explicit expressions for the $f,g,h$ functions. The relevant derivatives, to be plugged in (\ref{J}), read
\bea
\label{fMZ}
\frac{\partial{f}}{\partial M_Z}=
-\frac{M_Z}{\mu}\frac{1}{2R_\mu^2}=-\frac{M_Z}{\mu_{\rm low}}\frac{1}{2R_\mu}
\eea
\bea
\label{ht}
\frac{\partial{h}}{\partial t}=B_{\rm low}\ \frac{1-t^2}{t(1+t^2)} 
\eea
\bea
\label{gmt}
\frac{\partial{g}}{\partial m_t}=
\frac{E}{v\ s_\beta}\left(\frac{y}{y_{\rm low}} \right)^2 \ .
\eea
Let us comment briefly on these expressions. As mentioned above, eq.~(\ref{fMZ}) is essentially the 
fine-tuning factor $2\mu/(M_Z c_\mu)$ obtained in subsect.~2.1 [eq.~(\ref{pcmu})]. It penalizes
large scales for $\mu$. Eq.~(\ref{ht}) counts the volume conversion from $dB$ to $dt$ and it is proportional to a soft mass just for dimensional reasons. Note that this factor penalizes low scales. This is easy to understand looking at eq.~(\ref{B}): 
for a given interval in $\tan\beta$, the larger the values of the soft masses and $\mu$, the larger the corresponding interval in $B$ is. So larger $B$ is favoured. Note, however, that the size of the interval of $B$ relative to the value of $B$ itself (which is statistically meaningful) is essentially constant. Indeed, the $B$-factor in eq.~(\ref{ht}) will be cancelled in the pdf if one uses logarithmic flat priors for the soft terms, $p(B)\propto 1/B$. This reasoning is similar to that after eq.~(\ref{pcmu}). Finally, eq.~(\ref{gmt}) corresponds to eq.~(\ref{marg_y}) of our preliminar discussion. In particular, the $1/s_\beta$ factor corresponds to the same factor in
(\ref{marg_y}). 
 
The Jacobian of the transformation (\ref{change_3}) is given by the product of the three factors of eqs.~(\ref{fMZ}--\ref{gmt}),
\bea
\label{Jexpl}
J\ = \ \frac{1}{4}(g^2+g'^2)^{1/2}\left[\frac{E}{R_\mu^2}\right]\ \frac{B_{\rm low}}{\mu}
\frac{t^2-1}{t(1+t^2)} \left(\frac{y}{y_{\rm low}}\right)^2 s_\beta^{-1}\ .
\eea

In the previous derivation we have considered just the top Yukawa coupling in the change of variables (\ref{change_3}). Once the
others fermions are taken into account, the Jacobian gets a $s_\beta^{-1}$ factor for each 
$u-$type quark and a $c_\beta^{-1}$ factor for each $d-$type quark and charged lepton, as discussed in subsect.~2.2. 
Now, recall that the effective prior in the new variables is the product of $J$ by the initial prior, as expressed in eqs.~(\ref{pdf_new}, \ref{eff_prior}); so taking a logarithmically flat prior for the Yukawa couplings (i.e. $p(y_i)\propto y_i^{-1}$) the $s_\beta^{-1}, c_\beta^{-1}$ factors get cancelled
in the effective prior and the pdf. For the top Yukawa coupling (and sometimes for the bottom one) this cancellation still leaves a residual dependence on $\tan\beta$ since
$\displaystyle{\left(\frac{y}{y_{\rm low}}\right)^2 s_\beta^{-1}\times \frac{1}{y}\propto
\frac{y}{y_{\rm low}}}$, which through (\ref{yLH}) depends on $y$ itself and thus on $\tan\beta$. 

Therefore, the effective prior defined by eq.~(\ref{eff_prior}) takes the approximate form
\bea
\label{approx_eff_prior}
p_{\rm eff}(m_t, m, M, A, \tan\beta)\ \ \propto \   \left[\frac{E}{R_\mu^2}\right]\ 
\frac{y}{y_{\rm low}} \frac{t^2-1}{t(1+t^2)} 
\frac{B_{\rm low}}{\mu_0}  
p(m, M, A, B, \mu=\mu_0)\ 
.
\eea
The most basic priors for the initial variables are the flat and the logarithmic ones, i.e.
\bea
\label{priors}
p(m, M, A, B, \mu)\ =\ {\rm const.}\ , \;\;\;\;\;\;\;\;
p(m, M, A, B, \mu)\ \sim\ \frac{1}{mMAB\mu}\ 
.
\eea
Some comments are in order here. First, the normalization factors in (\ref{priors})
are determined by the integrated probability and thus depend on the bounds one establishes for the parameters. Since we are discussing here relative probabilities in the parameter space, they are not relevant at this stage, but they become more important when some parameters are marginalized. Second, as argued in subsect.~2.1, the logarithmic prior is physically sensible and is the one that can catch the intuition that fine-tunings are statistically unlikely. Actually, when plugged in (\ref{approx_eff_prior}), the logarithmic prior gives rise to the fine-tuning penalization $1/\mu^2\sim 1/c_\mu$.
However, the simple logarithmic prior of eq.~(\ref{approx_eff_prior}) is clearly too simple, since it cannot be normalized due to low-energy and high-energy divergences.
These are easily cured by taking reasonable upper and lower bounds on the parameters, e.g. $[10\ {\rm GeV},\ M_X]$. In fact, this choice can be refined. From the 1-loop RGE of the initial parameters, it is clear that very small values for $m,A,B$ are not radiatively stable, due to sizeable contributions proportional to the gaugino mass $M$. Therefore, it is not very sensible to assume that values of these parameters smaller than say 
${\cal O}(10^{-1} M)$ at precisely $M_X$ can have a particular statistical meaning.
Thus we can take flat priors at this region of small values. On the other hand, the  experimental lower bounds on the gluino, charginos and neutralinos imply that $M$ and $\mu$ cannot be smaller than ${\cal O}(100)$ GeV. 
\begin{figure}
\centering
\includegraphics[angle=0.,width=.46\linewidth]{jac-tb3-a0-mu+-fp.eps}\;\;\;\;\;\;\;\;
\includegraphics[angle=0,width=.46\linewidth]{jac-tb3-a0-mu+-logp.eps}\vspace{0.1cm}
\includegraphics[angle=0,width=.46\linewidth]{jac-tb10-a0-mu+-fp.eps}\;\;\;\;\;\;\;\;
\includegraphics[angle=0,width=.46\linewidth]{jac-tb10-a0-mu+-logp.eps}\vspace{0.1cm}
\includegraphics[angle=0,width=.46\linewidth]{jac-tb30-a0-mu+-fp.eps}\;\;\;\;\;\;\;\;
\includegraphics[angle=0,width=.46\linewidth]{jac-tb30-a0-mu+-logp.eps}\vspace{-0.5cm} 
\includegraphics[angle=0,width=0.7\linewidth]{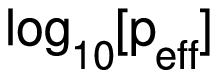}\vspace{-0.1cm}
\includegraphics[angle=0,width=.46\linewidth]{cbar-flat.eps}\hspace{1cm}
\includegraphics[angle=0,width=.46\linewidth]{cbar-log.eps}\vspace{0.2cm}

\caption{Values of the effective prior, $p_{\rm eff}$, in logarithmic units 
as defined in eq.~(\ref{eff_prior}) (up to a normalization constant), in the $\{m, M\}$ plane for $A=0$ and $\tan\beta=3$ (upper plots),  $\tan\beta=10$ (central plots), $\tan\beta=30$ (lower plots). The left and right plots correspond respectively to the two basic choices of priors (flat and logarithmically flat) discussed in eq.~(\ref{priors}) and below. See text for further details.
\label{fig:eff_prior}}
\end{figure}

In Fig.~2 we show the effective prior defined in (\ref{eff_prior}) and computed using eq.~(\ref{J}) [with the full one-loop expressions of eqs.~(\ref{mu}--\ref{y})] 
for the two priors discussed after eq.~(\ref{priors}), i.e. flat and logarithmically flat.
The plots show, up to a constant of proportionality, 
the effective priors in the 
$\{m,M\}$ plane (with constant $\tan\beta, A$) 
for some representative cases\footnote{The proportionality constant is simply the normalization constant of the initial prior, eqs.~(\ref{priors}), times the normalization constant of the Yukawa prior. However, these factors play no role in the exam of the relative probabilities of the points in the parameter space. Obviously, the absolute values of the left plots cannot be compared to those of the right plots, as they are affected by a different normalization constant. We prefer not to include these normalization factors, as they depend on the upper limit assumed for the soft terms, and do not shed any additional light on the relative probabilities inside the parameter space.}. We have assumed in the figures that the soft terms are initially given at 
the scale of gauge unification, $M_{\rm X}\sim 10^{16}\ {\rm GeV}$, as essentially happens in scenarios of gravity--mediated SUSY breaking, but of course our formulas are also applicable to e.g. gauge--mediated SUSY breaking scenarios.
The penalization of large scales is clear for the logarithmically flat case, as expected from our discussion.  The fact that using a logarithmic prior penalizes large values of the parameters could seem quite obvious. However, this is not so clear when one compares 
the integrated probability that the 
the parameters are within different ranges of scales. For instance, the logarithmic prior alone would give more probability to the $[100\ {\rm TeV}, M_X]$ range than to the $[100\ {\rm GeV}, 100\ {\rm TeV}]$ one. However, the presence of the mentioned fine-tuning factor, $1/\mu^2$, in the {\em effective} prior still penalizes the high-energy regions.

\begin{figure}
\centering
\includegraphics[angle=0,width=.45\linewidth]{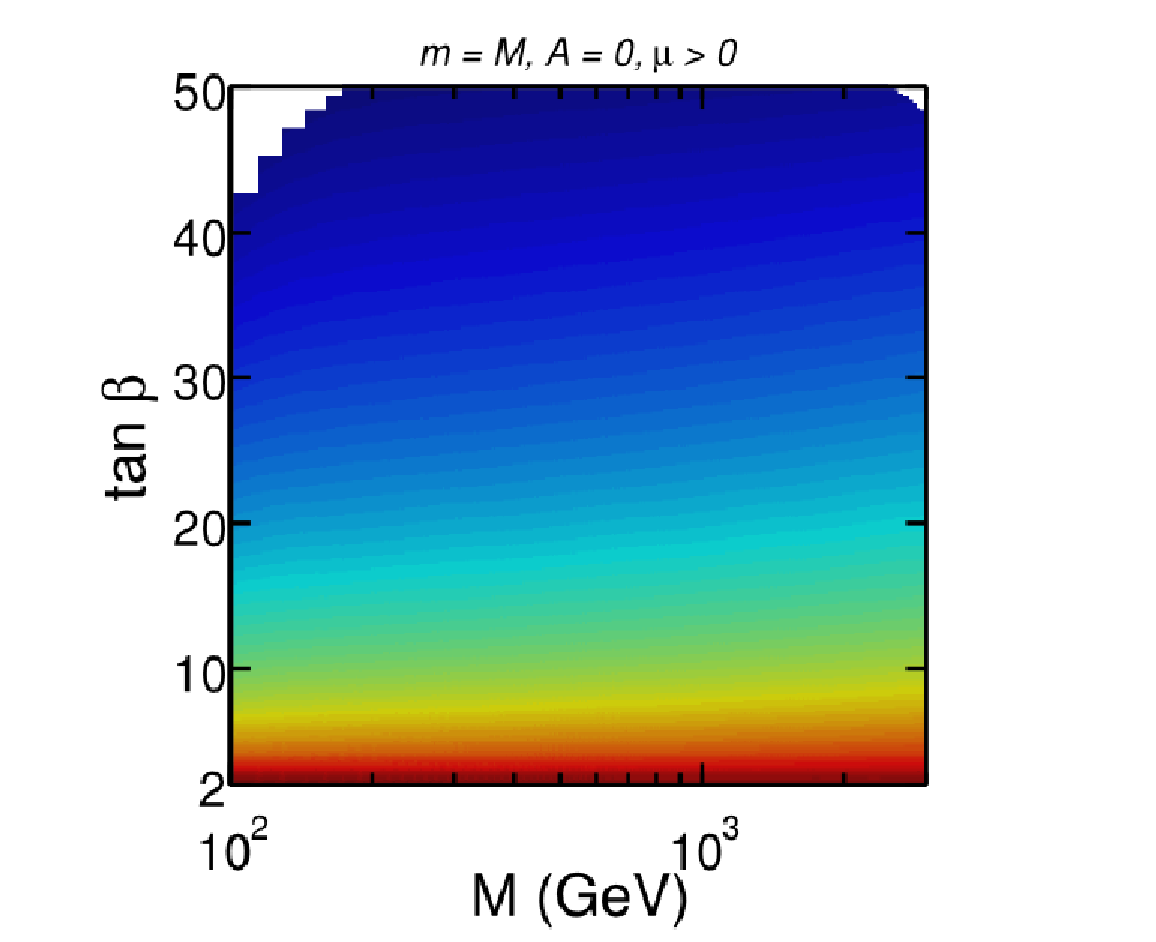}\;\;\;\;\;\;\;\;
\includegraphics[angle=0,width=.45\linewidth]{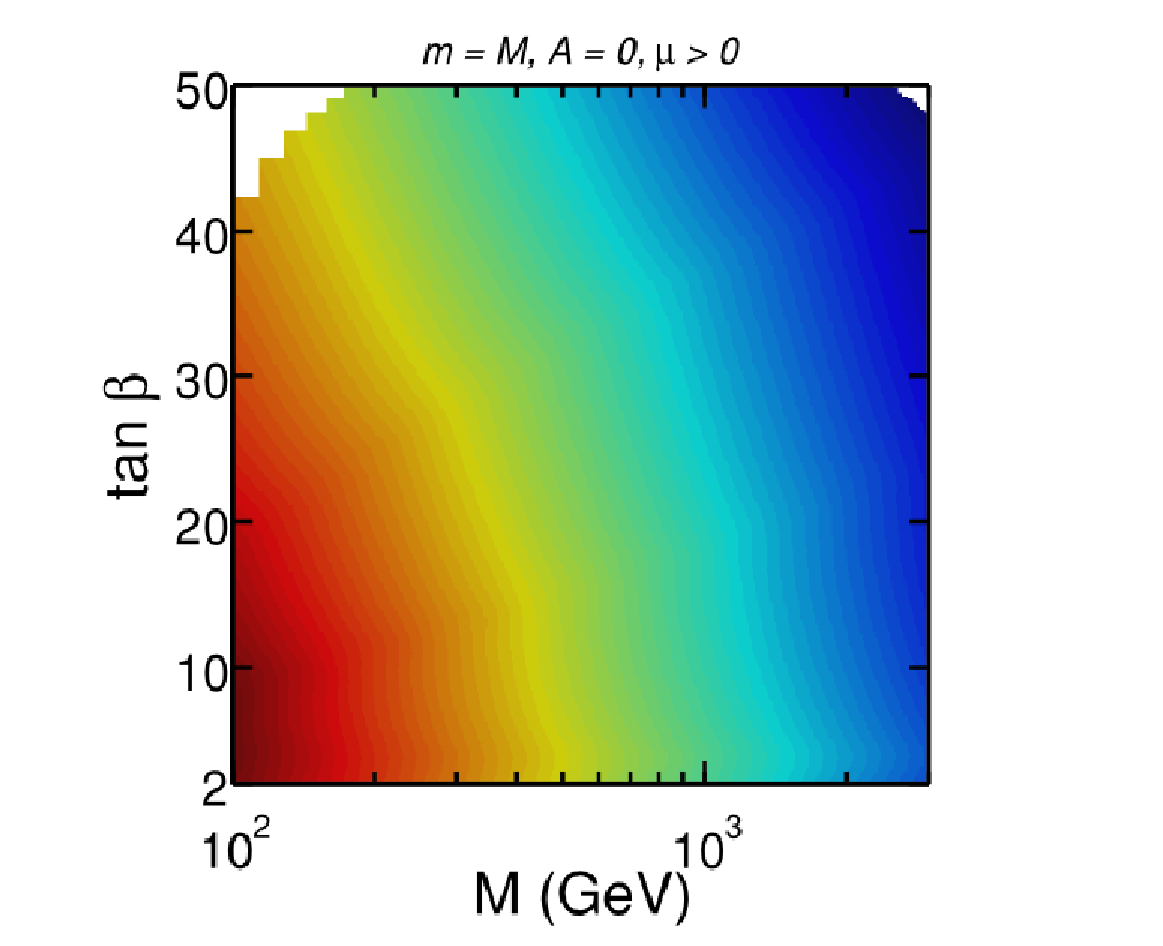}\vspace{0.2cm}
\includegraphics[angle=0,width=0.7\linewidth]{peff-label.eps}\vspace{-0.1cm}
\includegraphics[angle=0,width=.45\linewidth]{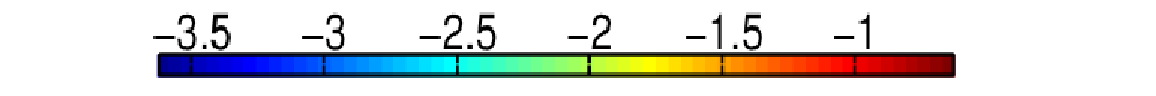}\;\;\;\;\;\;\;\;
\includegraphics[angle=0,width=.45\linewidth]{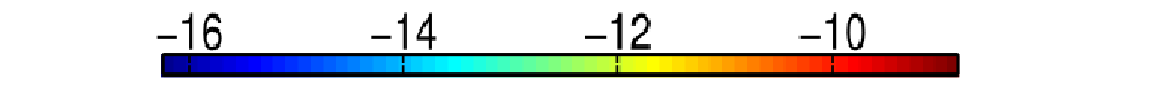}\vspace{0.2cm}

\caption{The same as Fig.~2, but in the $\{M,\tan\beta\}$ plane, for $A=0$, 
$m=M$.
\label{fig:eff_prior_m=M}}
\end{figure}

Fig.~3 is similar to Fig.~2, but showing now slices in the $\{M, \tan\beta\}$ plane (with the condition $m=M$).  The plots illustrate the
$\tan\beta$ dependence of the effective prior, which can be essentially extracted from the approximate expression (\ref{approx_eff_prior}). [Note that, besides the
explicit dependence, eq.~(\ref{approx_eff_prior}) contains an implicit dependence on $\tan\beta$ through the $R_\mu$, $B_{\rm low}$ and 
${y}/{y_{\rm low}}$ factors.]
We can appreciate from the plots that the prior probability decreases with $\tan\beta$.

The effective prior computed and shown in the figures corresponds to the last two factors of the pdf (\ref{pdf_new}). The first factor, i.e. the likelihood, carries the experimental information (fermion masses, electroweak precision tests, g-2 of the muon, dark matter constraints, etc.). Whatever experimental information (and thus likelihood) we may use, it will be always weighted by the same effective prior factor shown here.

In this section we have argued so far that the sensible initial choice of independent parameters of the MSSM is $\{g_i, y_t, m, M, A, B, \mu\}$, while for practical reasons it is most convenient to work with the set $\{g_i, m_t, m, M, A, \tan\beta, M_Z\}$ (and ${\rm sign}\mu$). $M_Z$ is eliminated from the analysis using its extremely sharp likelihood. 
The effective prior in the new variables is then given by eqs.~(\ref{eff_prior}, \ref{J}), for which we gave explicit approximate expressions in eqs.~(\ref{Jexpl}, \ref{approx_eff_prior}). 

It is interesting to wonder what would have been the result if one had insisted in taking directly $m_t$ as an initial (nuisance) variable, so that the transformation (\ref{change_3}) would have just involved 
$\{\mu,B\}\ \rightarrow\  \{M_Z,t\}$, as has been done e.g. in ref.~\cite{Allanach:2007qk}. As argued in subsect.~2.2, it is theoretically bizarre to take $m_t$ as a fundamental variable, instead of $y_t$. However, one may gain the bonus of almost no sensitivity to the prior in $m_t$, since this is essentially fixed by the experiment. This is true, but this procedure introduces extremely counter-intuitive contributions to the Jacobian, as we will see briefly. The new 2-variable Jacobian is given by
\bea
\label{J2}
J_2 = \left|\begin{array}{cc} 
\left.\frac{\partial\mu}{\partial M_Z}\right|_{t, m_t}
& \left.\frac{\partial\mu}{\partial t}\right|_{M_Z,m_t} \\
& \\
\left.\frac{\partial B}{\partial M_Z}\right|_{t,m_t}
& \left.\frac{\partial B}{\partial t}\right|_{M_Z,m_t} 
\end{array}\right|\ \ ,
\eea
where the subscripts emphasize which variables have to be kept frozen in the partial derivations. Now, using the definitions (\ref{muyB}), it is straightforward to obtain
\bea
\label{J2_expl}
J_2 =\ \frac{\partial f}{\partial M_Z}\ \frac{\partial h}{\partial t} 
\ +\ \frac{\partial g}{\partial M_Z}\left( \frac{\partial f}{\partial y}\ \frac{\partial h}{\partial t} -  \frac{\partial f}{\partial t}\ \frac{\partial h}{\partial y}  \right)
\ +\ \frac{\partial f}{\partial M_Z}\ \frac{\partial h}{\partial y}\ \frac{\partial g}{\partial t}\ .
\eea
It is amusing that this expression is more complicated than in the 3-variable case, eq.~(\ref{J}). 
This comes from the fact that the derivatives in (\ref{J2}) contain contributions coming from the dependence of $\mu$ and $B$ on $y$, which is in turn a function of $t$ and $M_Z$, eq.~(\ref{y}).
These contributions were cancelled inside the 3-variable Jacobian thanks to the third row in the matrix of eq.~(\ref{J}), but they are not cancelled here and give rise to the second and third terms in eq.~(\ref{J2_expl}).
Note that the first term in (\ref{J2_expl}) is similar to the 3-variable Jacobian given by 
eq.~(\ref{J}), whose physical significance (including the information about fine-tuning) was discussed after eq.~(\ref{gmt}).
This term goes parametrically as $B/\mu$ and was the only one quoted in ref.~\cite{Allanach:2007qk}, thus the resemblance of their result to our approximate expression
(\ref{Jexpl}), except for the RG and $s_\beta^{-1}$ factors. 
However the second term goes parametrically as $Bm^2/\mu M_Z^2$, and thus is much more important for large soft terms, which then become strongly favoured (contrary to the intuitive expectatives). Therefore there is no reason to have ignored such term.
In consequence, the expressions used in ref.~\cite{Allanach:2007qk} are much closer to using $y_t$ as a fundamental variable with logarithmically flat prior than to using $m_t$.

Let us finish this section by using the approximate expressions discussed above to give, as advanced at the end of subsect.~2.1, an approximate expression for the fine-tuning parameter $c_\mu$. Recall that this parameter was defined as
\bea
\label{cmu}
c_\mu = \left|\frac{\partial \ln M_Z^2}{\partial \ln \mu}\right|_{y,B},
\eea
where the subscript indicates that the partial derivative must be performed at $y,B$ constant. Using eqs.~(\ref{muyB}), $c_\mu$ can be written as
\bea
\label{cmu2}
c_\mu = \frac{2\mu}{M_Z}
 \left(\frac{\partial f}{\partial M_Z}\right)^{-1}\left[1\ +\ \frac{\partial f}{\partial t}
\ \frac{\partial h}{\partial \mu}\ \left(\frac{\partial h}{\partial t}\right)^{-1}\right] \;,
\eea
where the right hand side has to be understood in absolute value. As above, 
using
(\ref{muBLH}) and (\ref{yLH}) we obtain explicit approximated expressions for the $f,g,h$ functions. Then eq.~(\ref{cmu2}) reads
\bea
\label{cmu3}
c_\mu = 4 R_\mu^2\frac{\mu^2}{M_Z^2} \left[
1\ -\ \frac{t^2(1+t^2)}{(t^2-1)^3} \frac{m_{H_1}^2 - m_{H_2}^2}
{B_{\rm low}\mu_{\rm low}}
\left( \frac{B_{\rm low}}{\mu_{\rm low}} - \frac{4t^2}{1+t^2}\right)
\right]\ .
\eea
Note that the combination $m_{H_1}^2 - m_{H_2}^2$ can be easily written in terms of
$B,\mu$ using eqs.~(\ref{mu}, \ref{B}).

\section{Conclusions}

The start of LHC has motivated an effort
to determine the relative probability of the different regions of the MSSM parameter space, taking into account the present (theoretical and experimental) wisdom about the model. These attempts are often called ``LHC forecasts" \cite{Allanach:2005kz, Allanach:2006jc, deAustri:2006pe, Allanach:2007qk, Buchmueller:2008qe, Trotta:2008bp, Ellis:2008di}. The central equation to extract this valuable information is 
the fundamental Bayesian relation
\bea
\label{Bayes2}
p(s,m,M,A,B,\mu|{\rm data})\ \propto \ {\cal L}(s,m,M,A,B,\mu)\ p(s,m,M,A,B,\mu)\ ,
\eea
which gives this probability in terms of the usual experimental likelihood, ${\cal L}$, and the prior $p(s,m,M,A,B,\mu)$,  i.e. the ``theoretical" probability density assigned a priory to points in the space spanned by the MSSM parameters $\{m,M,A,B,\mu\}$ and the SM-like ones ($s$). 

Since the present experimental data are not powerful enough to select a small region of the MSSM parameter space, the choice of a judicious prior becomes most relevant. Indeed, ignoring this amounts to an implicit choice for the prior (which is not always sensible).
On the other hand, it is common lore that the parameters of the MSSM, $\{m,M,A,B,\mu\}$, should not be far from the electroweak scale in order to avoid unnatural fine-tunings to obtain the correct scale of the electroweak breaking. Previous studies have attempted to incorporate this reasonable intuition to the Bayesian approach, by choosing a prior that counted (more or less explicitly) a conventional measure of the fine-tuning, typically the Barbieri-Giudice parameter, $c$, defined in eq.~(\ref{BG}).

However, though reasonable, these kinds of proposals are rather arbitrary, as the very measure of the fine-tuning is. On the other hand, since the naturalness arguments are deep down statistical arguments, one might expect that an effective penalization of fine-tunings
should arise from the Bayesian analysis itself. One of the main results of this paper has been to show that this is really so: using the fact that the likelihood associated to the experimental $M_Z$ is essentially a Dirac delta, $\sim \delta(M_Z-M_Z^{\rm exp})$, one can easily marginalize the $\mu$-parameter (i.e. integrate the density of probability in this variable). Then one gets an effective prior for the remaining parameters
\bea
\label{c_factor2}
p_{\rm eff}(s, m, M, A, B)\
=\ 2\ \frac{\mu_0}{M_Z}\ \frac{1}{c_\mu}\
p(s, m, M, A, B,\mu_0)\ ,
\eea
which exhibits the fine-tuning penalization. ($\mu_0$ is the value of $\mu$ that reproduces the experimental 
$M_Z$ for the given values of $\{s, m, M, A, B\}$.)
Of course this effective prior has to be combined with the experimental likelihood, except the part associated to the $Z$ mass. The initial prior, $p(s, m, M, A, B,\mu)$, can be taken as flat or (preferably) logarithmically flat, as usual. 
We find very satisfactory that precisely the usual parameter to quantify the degree of fine-tuning emerges in the Bayesian approach ``spontaneously", not upon subjective assumptions, especially taking into account that there has been much discussion in the literature about its significance and suitability. We have completed this analysis by giving an explicit and quite accurate expression for $c_\mu$, see eq.~(\ref{cmu3}).

Our second result concerns the treatment of the Yukawa couplings. 
In previous Bayesian analyses the Yukawas were essentially taken as needed to reproduce the experimental fermion masses, within uncertainties. However, unlike the pure SM, in the MSSM the Yukawa couplings are not in one-to-one correspondence to the quark and lepton masses: they depend also on the value of $\tan\beta$, which is a derived quantity that takes different values at different points of the MSSM parameter space. This means that two viable MSSM models (with the same fermion masses) will have in general very different values of the Yukawa couplings, and thus the theoretical prior, $p(y)$, will play a relevant and non-ignorable role in evaluating their relative probability. Any Bayesian analysis of the MSSM amounts to an explicit or implicit assumption about the prior in the Yukawa couplings. We have made explicit the dependence of the results on such prior and shown that the easiest and usual practice of taking the Yukawas ``as required", 
approximately corresponds to taking logarithmically flat priors in the Yukawa couplings, which on the other hand is not an unreasonable choice at all.

Finally we have repeated this analysis, using a more efficient set of variables to scan the MSSM parameter space. Besides trading $\mu$ by $M_Z$ and the Yukawa couplings (in particular the top one) by the fermion masses, it is known that trading $B$ by $\tan\beta$ is highly advantageous. Following similar steps one can arrive to an effective prior in the new parameters:
\bea
\label{eff_prior2}
p_{\rm eff}(g_i, m_t, m, M, A, \tan\beta)\  \equiv\ 
J|_{\mu=\mu_0}\  p(g_i, y_t, m, M, A, B, \mu=\mu_0)\ ,
\eea
where $J$ is the Jacobian of the transformation
\bea
\label{change_32}
\{\mu,y_t,B\}\ \rightarrow\  \{M_Z,m_t,t\},\;\;\;\;\;\;t\equiv\tan\beta
\eea
($M_Z$ does not appear in the right hand side of (\ref{eff_prior2}) since it is marginalized as explained above.) Note that still  the initial choice of independent parameters is $\{y_t, m, M, A, B, \mu\}$ (on which the initial priors are defined).
It is the change of variables plus the marginalization of $M_Z$ what leads to the above effective prior. We have calculated $J$ both numerically and analytically
(in an approximate but quite accurate fashion). The relevant formulas are eqs.~(\ref{J}) and(\ref{Jexpl}). The last expression is very handful and leads to the effective prior given in eq.~(\ref{approx_eff_prior}). 
Whatever experimental information (and thus likelihood) one may use, it will be always weighted by the same effective prior factor calculated (and shown in plots for illustrative cases) here.

We have also discussed the results in comparison with other approaches in the literature, arguing that the present one is conceptually more satisfactory.

\section*{Acknowledgements}
We thank D. Garc\'{\i}a-Cerde\~no and R. Trotta for interesting discussions and suggestions. This work has been partially supported by the MICINN, Spain, under contract FPA 2007--60252, the Comunidad de Madrid through Proyecto HEPHACOS S-0505/ESP--0346, and by the European Union through the UniverseNet (MRTN--CT--2006--035863).
M.~E. Cabrera acknowledges the 
financial support of the CSIC through a predoctoral research grant (JAEPre 07 00020).
R. Ruiz de Austri is supported by the project PARSIFAL (FPA2007-60323) of the 
Ministerio de Educaci\'{o}n y Ciencia of Spain .
The use of the ciclope cluster of the IFT-UAM/CSIC is also acknowledged.

\end{document}